\documentclass[artical]{IEEEtran}

\usepackage{amsmath}

\ifCLASSINFOpdf
\usepackage[pdftex]{graphicx}
  
\else
 
\fi

\begin{document}
\title{A Memristive Model Compatible with Triplet Rule for Spike-Timing-Dependent-Plasticity}

\author{Weiran Cai,~\IEEEmembership{Student Member, IEEE}, Ronald Tetzlaff,~\IEEEmembership{Senior Member, IEEE},
        Frank Ellinger,~\IEEEmembership{Senior Member, IEEE}
\thanks{Weiran Cai and Frank Ellinger are with the Chair for Circuit Design and Network Theory, and Ronald Tetzlaff is with the Chair for Fundamentals of Electronics, Faculty of Electrical Engineering and Information Technology, Technische Universit\"at Dresden, Helmholtzstrasse 18, Barkhausenbau, 01069 Dresden, Germany. e-mail: weiran.cai@tu-dresden.de.}
}

\maketitle

\begin{abstract}
In this paper, we propose an extended version of the memristive STDP model, which is one of the most important and exciting recent discoveries in neuromorphic engineering. The proposed model aims to claim compatibility with another importent STDP rule beyond the pair-based rule, known as the triplet STDP rule. This is an extension of the asynchronous memristive model of Linares-Barranco, et al., capable of explaining the pair-based rule based on the analogy of the synapse to the memristor. The proposed new model is compatible with both the pair-baed and triplet-based rule, by assuming a mechanism of variable thresholds adapting to synaptic potentiation and depression. The dynamical process is governed by ordinary differential equations. The model is an expression of Froemke's principle of suppression for triplet rules and reveals a similar time dependence with that in the suppression STDP model.   
\end{abstract}


\IEEEpeerreviewmaketitle

\section{Introduction}
Since decades, the study on nervous systems has aroused great interest in researchers that span across mulitple fields, growing out far beyond a single research branch to a complex scientific territory of mutually inspiring disciplines involving biology, mathematics, physics, electrical engineering and computer science. Biological discoveries have been inspiring engineers in the sense of realizing intellectual systems and networks, while concepts in arificial systems are also providing new insights in understanding biological nervous systems. Recently, an important discovery was made on the explanation of the spike-timing-dependent-plasticity (STDP) in the neuronal communications based on a memristor model \cite{Snid}-\cite{Jo}. The STDP rule is one of the fundamental properties in the neurons, which permits adaptation of developing nervous systems to the surrounding environment \cite{Gers1}-\cite{Abbo}, whereas the memristor, on the other side, is considered as a fourth fundamental electrical element, which was originally proposed by L. Chua \cite{Chua1}\cite{Chua2} and recently implemented in material at a nano-scale \cite{Stru}. The researchers point out in \cite{Lina3} that the synapse between a pair of communicating neurons has similar function of a memristor, attributed to the long-term memory of weight changes in the device. This connection makes it possible to realize an intellectual system by the means of highly integrated nano-devices, in the direction of which several research groups have proposed realizable schemes, leading to either synchronous or asynchronous systems. Most notably, in understanding the long-term plasticity of neurons biologically, this mechanistic model also provides a completely new way by uniquely taking into account two important properties of the neurons: the specific waveforms of the action potentials and a synaptic threshold. These properties makes the model different from the original behavioral model of Gerstner and Pfister, esp., in explaining the STDP update function, which is imposed in an ad hoc manner in the original model \cite{Kemp}.           

The pair protocol, consisting of a pair of pre- and post-synaptic spikes, is the basic rule in STDP. However, the memristive STDP model, like many other mechanistic models, has direct conflicts in explaining more complicated signal patterns such as the triplet rule. In \cite{Froe}, Froemke and Dan conducted a series of experiments on the triplet protocol in pyramidal neurons, which evaluates the synaptic modification involving three action potentials (single pre-synaptic and dual post-synaptic spikes comprising the 1/2 case, and dual pre-synaptic and single post-synaptic comprising the 2/1 case), and concluded their observations as: \textit{"The first spike pair played a dominant role in synaptic modification."} This is unexplainable either by the original pair-based model of Gerstner or by the memristive model, because these models assume that the two spike pairs contribute independently to the synaptic modification. In fact, the triplet protocol is considered as another important criteria for a STDP model, in parallel with the pair protocol \cite{Froe}-\cite{Wang}. In order to be compatible with both the pair and triplet rules, modifications have been made based on the existing pair-based models. Froemke and Dan proposed a suppression model by assuming a spike efficacy as an extra indicator based on the original STDP model, which is suppressed by preceding spike in each neuron \cite{Froe}. Pfister and Gerstner rather abandoned the basic assumption of paired spikes as the fundamental unit, and proposed a model considering that both pair and triplet are fundamental units to contribute to the synaptic modification \cite{Pfis}. The memristive STDP model, though explained successfully the pair-based STDP rule, is also confronting the triplet problem, which needs a modified version to be compatible with both pair and triplet protocols. In this paper, we exploit such possibility by proposing a memristive STDP model based on the mechanism of variable thresholds adapting to preceding synaptic modification. The modified memristive model, reserving the ansatz of paired spikes being the fundamental contributing unit, can explain the triplet rule in STDP by introducing suppression on a temporal base, which is a memristive realization of Froemke's principle. In the following sections, we will first briefly review the pair-based memristive STDP model and the confronting triplet problem. The memristive STDP model with adaptive thresholds will be proposed in Sec. III as a scheme of compatiblity with the triplet rule problem under the principle of suppression. In Sec. IV, we will point out potential biological relation to the process, and generalize the model to an extended form. A conclusion will be made in Sec. V.

\section{Pair-Based Memristive STDP Model and Triplet Rule}
\begin{figure}[!t]
\centering 
\includegraphics[width=8.4cm]{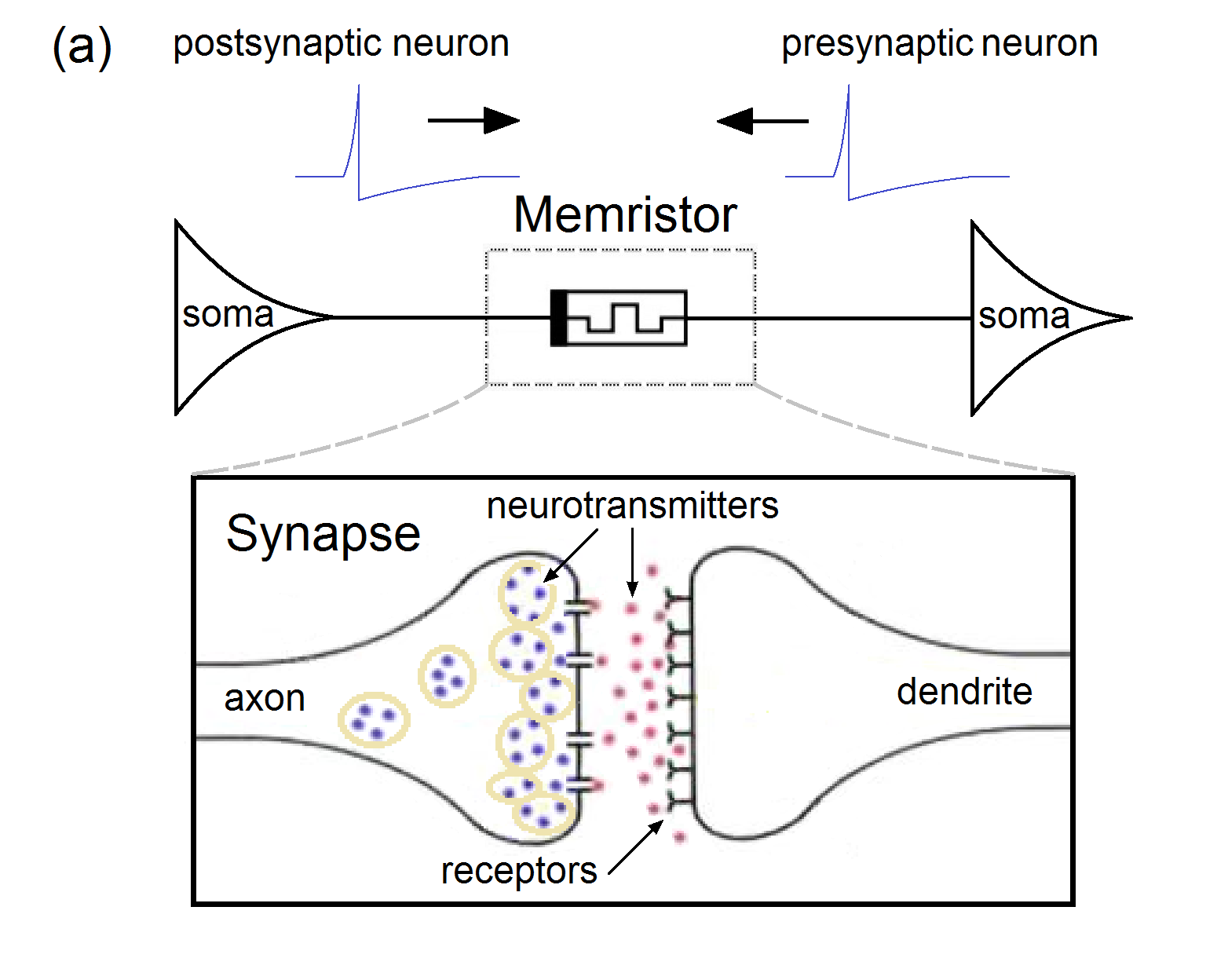}
\includegraphics[width=7.5cm]{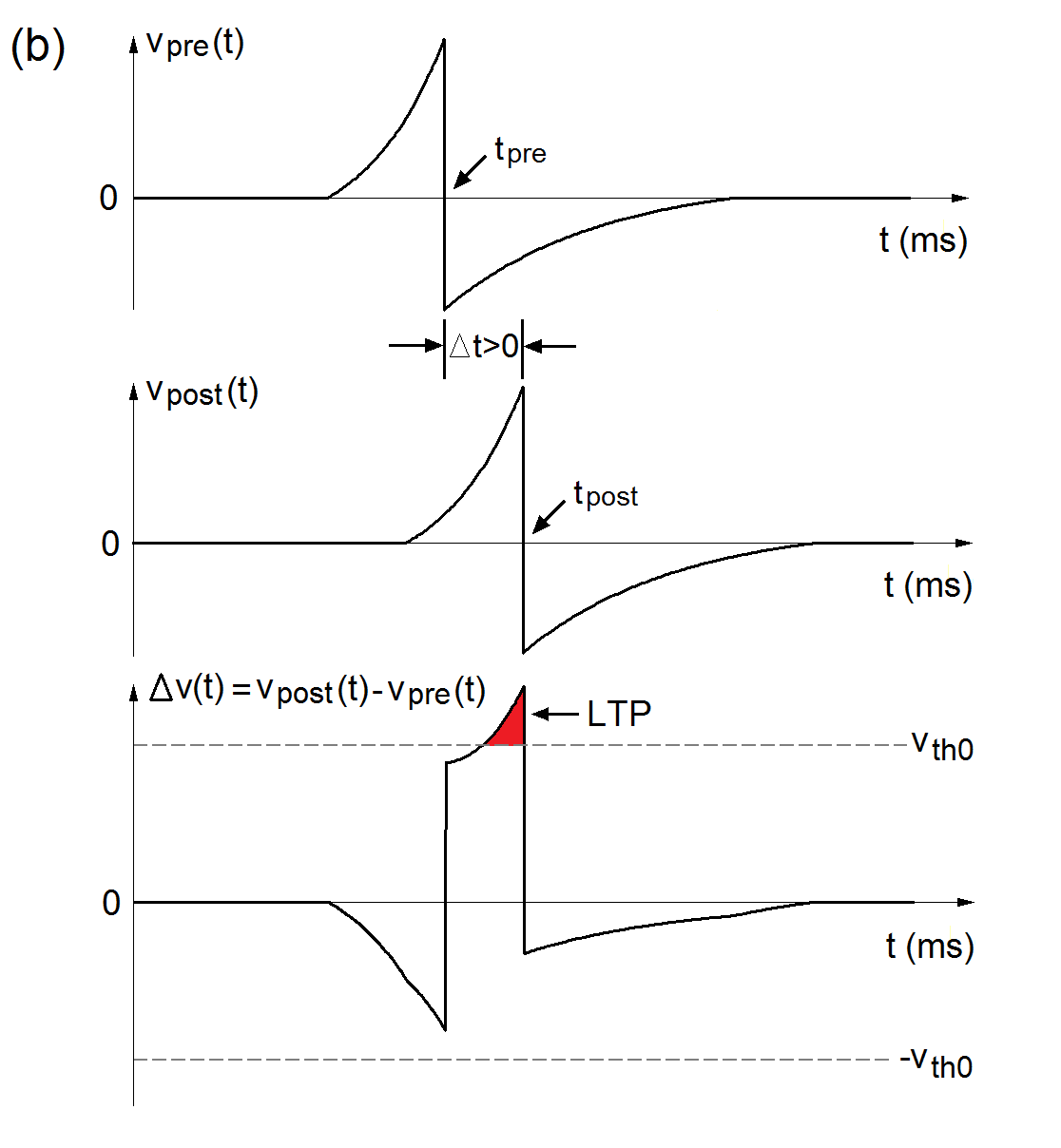}
\includegraphics[width=8.0cm]{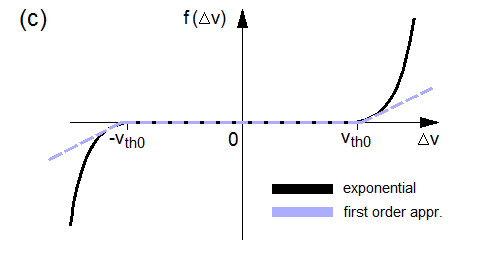}
\caption{The memristive pair-based STDP model. Plots are modified from \cite{Lina3}. (a) The analogy between a synapse and a memristor. (b) Demonstration of the pair-based model for the pre-post case (LTP). (c) The piecewise rate function $f$ of the synaptic modification and its first order approximation for small $\Delta v \pm v_{th0}$.
}\label{lina} 
\end{figure}

The STDP is a causality-based Hebbian rule with extension to both potentiation and depression, which features a critical role of relative timing of pre- and post-synaptic action potentials in the change of the synaptic efficacy \cite{Gers1}\cite{Hebb}. The causality is revealed in its principle that, within a certain time window involving a pair of spikes, pre-synaptic spike proceeding post-synaptic spike produces long term potentiation (LTP, positive change) in the synaptic efficacy, while post-synaptic spike proceeding pre-synaptic spike produces long term depression (LTD, negative change). This is often referred as the pair-based STDP rule. The memristive STDP models are established based on the analogy of long-term memory of the conductance (or indirectly, the weight) of the memristor to the efficacy of synapses. We here consider the asynchronous memristive STDP model proposed by Linares, et al. \cite{Lina3}. As demonstrated in Fig. \ref{lina}, the model assumes that the funtionality of a synapse can be modelled as a memristor with two thresholds. The modification of the conductance (the reciprocal of the memristance), corresponding to the synaptic efficacy, is governed by a function of the weight of the memristance (supposing that the memristance is linearly controlled by the weight $R(w)=k_R (w+w_0)$) and the voltage difference of the pre- and post-synaptic action potentials over the thresholds:    
\begin{equation}\label{dwdt_ful}
\frac{dw}{d t}=-f(w,\Delta v(t))
\end{equation}
where $\Delta v(t)$ is the difference between the membrane potentials
\begin{equation}\label{delta}
\Delta v(t)=v_{pos}(t)-v_{pre}(t)
\end{equation}    
The minus sign in Eq. (\ref{dwdt_ful}) is for excitatory synapses. In \cite{Lina3}, the function $f$ is chosen to be an exponential function solely on the voltage difference
\small\begin{eqnarray}\label{f}
  f(\Delta v) = \left\{ 
  \begin{array}{l l}
    I_0 \textrm{sign}(\Delta v) \left(\textrm{e}^{|\Delta v|/v_0}-\textrm{e}^{v_{th}/v_0}\right) &  \text{if } |\Delta v|>v_{th}\\
    0 &  \text{otherwise}\nonumber\\
  \end{array} \right.\\
\end{eqnarray}\normalsize
In this manner, the STDP update function can be predicted to be in accordance with experimental records \cite{Lina3}\cite{Bi}. This mechanism is based on paired spikes. 

\begin{figure}[!t]
\centering 
\includegraphics[width=8.2cm]{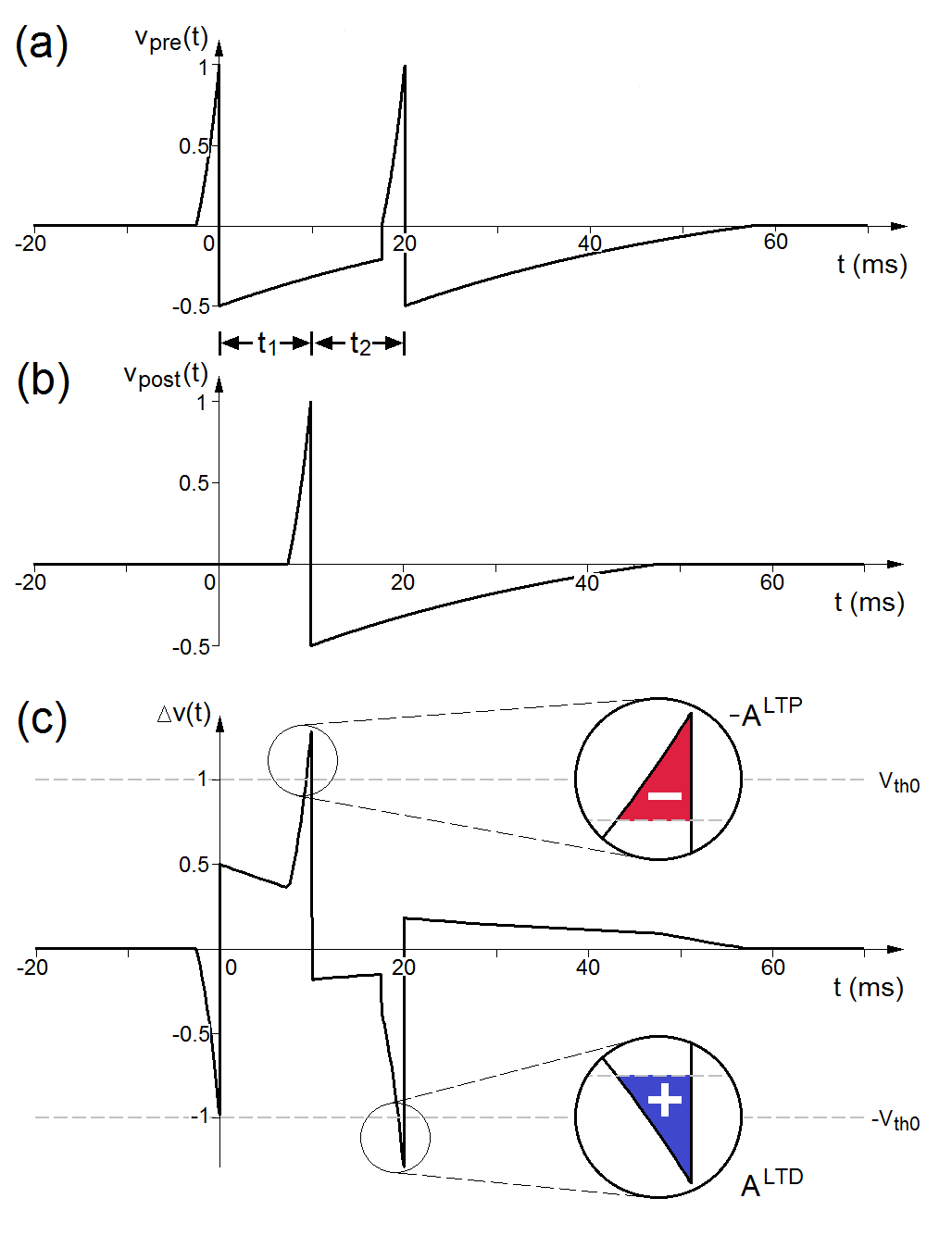}
\caption{The synaptic modification predicted by the pair-based memristive STDP model (the pre-post-pre case). (a) pre-synaptic spikes $v_{pre}(t)$: it is assumed that the first spike turns to the second spike abruptly at the presence of the second spike; (b) post-synaptic spike $v_{post}(t)$; (c) the voltage difference function $\Delta v(t)=v_{post}(t)-v_{pre}(t)$. For the pair-based model, the areas $A^{LTP}$ and $A^{LTD}$ are referred to the fixed thresholds $\pm v_{th0}$, respectively. The parameters for the spike function $spk(t)$ are chosen as follows: $A_{mp}^+=1$, $A_{ mp}^-=0.5$, $t_{ail}^+=2.5$ ms, $t_{ail}^-=37.5$ ms, $\tau_p=3$ ms, $\tau_m=40$ ms, the time intervals: $t_1=10$ ms, $t_2=10$ ms, and the threshold: $v_{th0}=1$.  
}\label{trip} 
\end{figure}

It is notable that the memristance is additive with such choice of function $f$ (whereas the synaptic efficacy is multiplicative). This implies that the total change of memristance is simply the sum of independent changes in all time intervals, with each consisting change depending only on $\Delta v(t)$, but not on the actual weight $w$ (or the memristance $R$)
\begin{equation}\label{delta_R}
\Delta R(t_0\rightarrow t_n)=-k_R\int_{t_0}^{t_n} f(\Delta v(t))dt=\sum_{k=1}^n\Delta R(t_{k-1}\rightarrow t_k)
\end{equation}
Its geometric implication can be seen intuitively. To the first order of Eq. (\ref{f}), the function $f(\Delta v)$ is a piecewise linear function of $\Delta v$ (the dashed line in Fig. \ref{lina}(c))
\begin{equation}\label{dwdt_lin}
f(\Delta v(t))=\frac{I_0}{v_0}\left([\Delta v(t)-v_{th}]_+ + [\Delta v(t)+v_{th}]_- \right)
\end{equation}
Hence, for small extra voltage over the thresholds $[\Delta v(t)-v_{th}]_+$ and $[\Delta v(t)+v_{th}]_-$ compared to the thresholds, the total change of memristance can be viewed as the net area enveloped by the curve $\Delta v(t)$ and the thresholds, multiplied by a factor $k_R I_0/v_0$ (see Fig. \ref{trip}(c))
\begin{equation}\label{delta_R1}
\Delta R(t_0\rightarrow t_n) \propto A^{LTD} - A^{LTP} 
\end{equation}

With the property of additivity, let us consider the synaptic modification corresponding to a triplet of spikes, involving two pre-synaptic and one post-synpatic spikes (the 2/1 case). According to the pair-based STDP rule and the additivity, the total change of the memristance is the sum of an LTP produced by the pre-post pair and an LTD produced by the post-pre pair (see Fig. \ref{trip}). Especially, when the interval between the first pair is equal to that of the second pair, the netto change of memristance sums to zero. However, this prediction is not in accordance with the triplet rule observed from experiments. In \cite{Froe}, Froemke and Dan concluded their experimental results that for triplet spikes, the first spike pair plays a dominant role in synaptic modification, or more specifically, the prediction based on the interval of the first pair ($t_1$) is better than that based on both intervals of the two pairs ($t_1$ and $t_2$) combined. It indicated that the two spike pairs do not contribute independently to synaptic modification, and the contribution of the second pair is strongly suppressed by the presence of the preceding pre-synaptic spike. The synaptic modification induced by the triplet, however, can be largely predicted by the second pair only when the first time interval is sufficiently large. This is the principle of suppression for the triplet STDP rule. For the pre-post-pre case, the synaptic modification should mainly express as LTP, and especially, should be nonzero for the case of equal intervals ($t_1 = t_2$). This discordance implies that the pair-based memristive model needs a modification to be compatible with the triplet STDP rule.

\section{Memristive STDP Model with Adaptive Thresholds}
\begin{figure}[!t]
\centering 
\includegraphics[width=7.8cm]{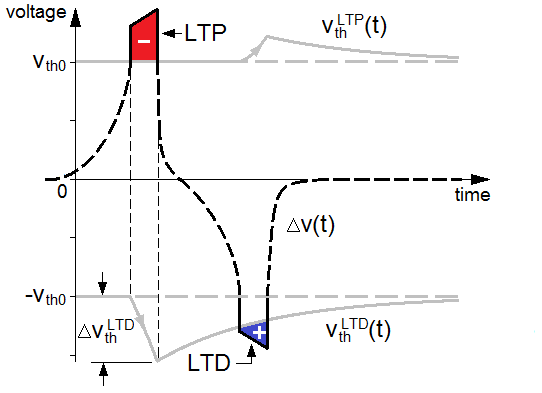}
\caption{The mechanism of the memristive STDP model with adaptive thresholds. The LTD area is suppressed by the increased threshold $v_{th}^{LTD}(t)$, which changes proportionally to the extent of the preceding LTP and returns exponentially to the resting value $-v_{th0}$. The suppression hence depends on the preceding LTP and the time interval between the LTD and the LTP. The function $\Delta v(t)$ is arbitrarily chosen for explanation.  
}\label{mech} 
\end{figure}

The principle of suppression for the triplet rule can be realized in the memristive STDP model by introducing a mechanism of adaptive thresholds. In this model, the thresholds are no longer rigid, but can vary according to the synaptic modification. We denote the positive threshold as the LTP threshold $v_{th}^{LTP}(t)$, and the negative threshold as the LTD threshold $v_{th}^{LTD}(t)$. It is supposed that the LTD threshold can rise by the presence of an LTP, while the LTP threshold can rise by the presence of an LTD. Both thresholds decrease exponentially to the resting values in the absence of LTP or LTD. The synaptic modification is then controlled not only by $\Delta v(t)$ but also by the actual value of the threshold $v_{th}^{LTP}(t)$ or $v_{th}^{LTD}(t)$. In this way, an LTP suppresses the following LTD to a certain extent by raising the LTD threshold, while an LTD also suppresses the following LTP in the same manner, as demonstrated in Fig. \ref{mech}. This process can be described by the following ordinary differential equation pair: 
\begin{eqnarray}
 \frac{dv_{th}^{LTP}}{d t}=\frac 1 {\tau_{th}^{LTP}} \left(-v_{th}^{LTP}(t) + v_{th0}\right) + k_{PD}\left[\frac{dw}{dt}\right]_+ \label{dvth_sim1}\\
 \frac{dv_{th}^{LTD}}{d t}=\frac 1 {\tau_{th}^{LTD}} \left(-v_{th}^{LTD}(t) - v_{th0}\right) + k_{DP}\left[\frac{dw}
{dt}\right]_- \label{dvth_sim2}
\end{eqnarray}
with the update of the synaptic weight
\begin{equation}\label{dwdt}
\tau_w \frac{dw}{d t}=-[\Delta v(t)-v_{th}^{LTP}(t)]_+ - [\Delta v(t)-v_{th}^{LTD}(t)]_-
\end{equation}
where $\pm v_{th0}$ denotes the resting LTP and LTD thresholds, respectively; $\tau_{th}^{LTP}$ and $\tau_{th}^{LTD}$ are the decreasing rates for the two thresholds. The changing rates of the thresholds are supposed to be linearly dependent on $dw/dt$ with the factors $ k_{PD}$ and $k_{DP}$, so that, by integration, the change in $v_{th}^{LTP}$ or $v_{th}^{LTD}$ is proportional to the change of $w$, i. e., $\Delta v_{th}^{LTP(D)}\propto \Delta w$, though generally it is not limited to linear dependence. For simplicity and geometric intuition, a linear synaptic update equation is also used in Eq. (\ref{dwdt}). 

\begin{figure}[!t]
\centering 
\includegraphics[width=8.6cm]{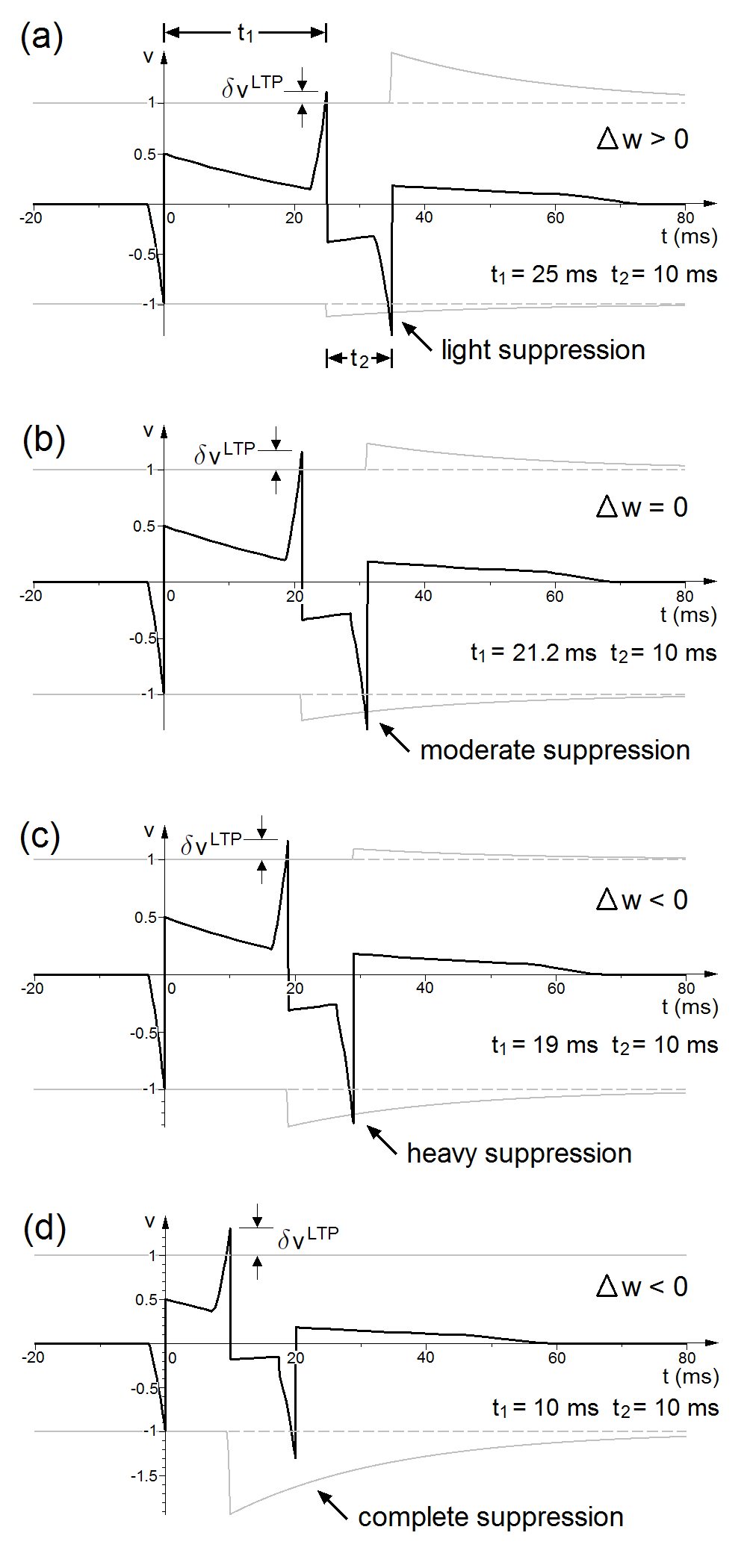}
\caption{Suppression by the adaptive thresholds in the 2/1 triplet spikes for different time intervals. $t_2$ is fixed at 10 ms, while $t_1$ is chosen at 25 ms (a), 21.2 ms (b), 19 ms (c) and 10 ms (d), respectively. Black lines: $\Delta v(t)$; grey solid: $v_{th}^{LTP}(t)$ and $v_{th}^{LTD}(t)$; grey dashed: $\pm v_{th0}$. Four typical suppression levels are shown: light, modest, heavy and complete suppression, which are produced by extra voltages ordered as $\delta v_{(a)}^{LTP}<\delta v_{(b)}^{LTP}<\delta v_{(c)}^{LTP}<\delta v_{(d)}^{LTP}$. The net synaptic modification is depression for (a), null for (b) and potentiation for (c) and (d). The parameters are chosen as: $\tau_{th}^{LTP}=\tau_{th}^{LTD}=25$ ms, $k_{PD}=k_{DP}=10$ ms$^-1$,The simulations are made with MAPLE v.13. 
}\label{syn} 
\end{figure}

\begin{figure}[!t]
\centering 
\includegraphics[width=9.6cm]{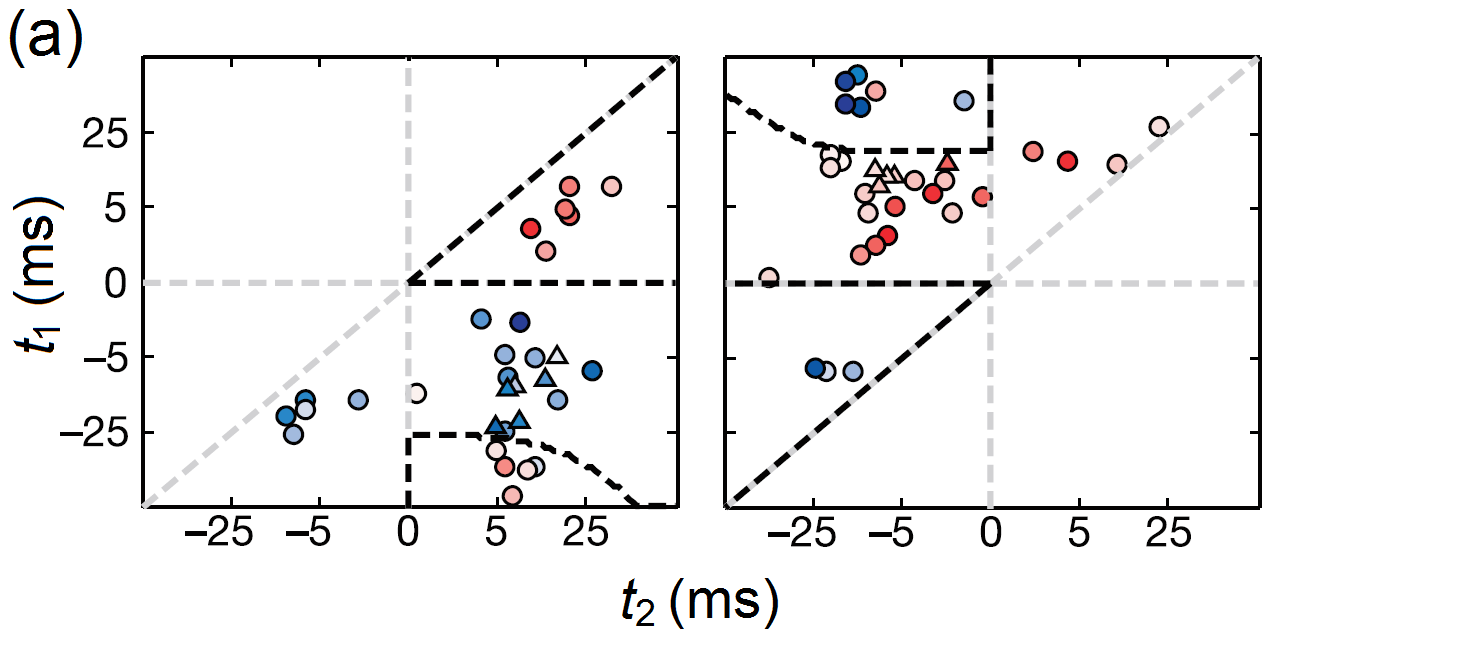}
\includegraphics[width=9.1cm]{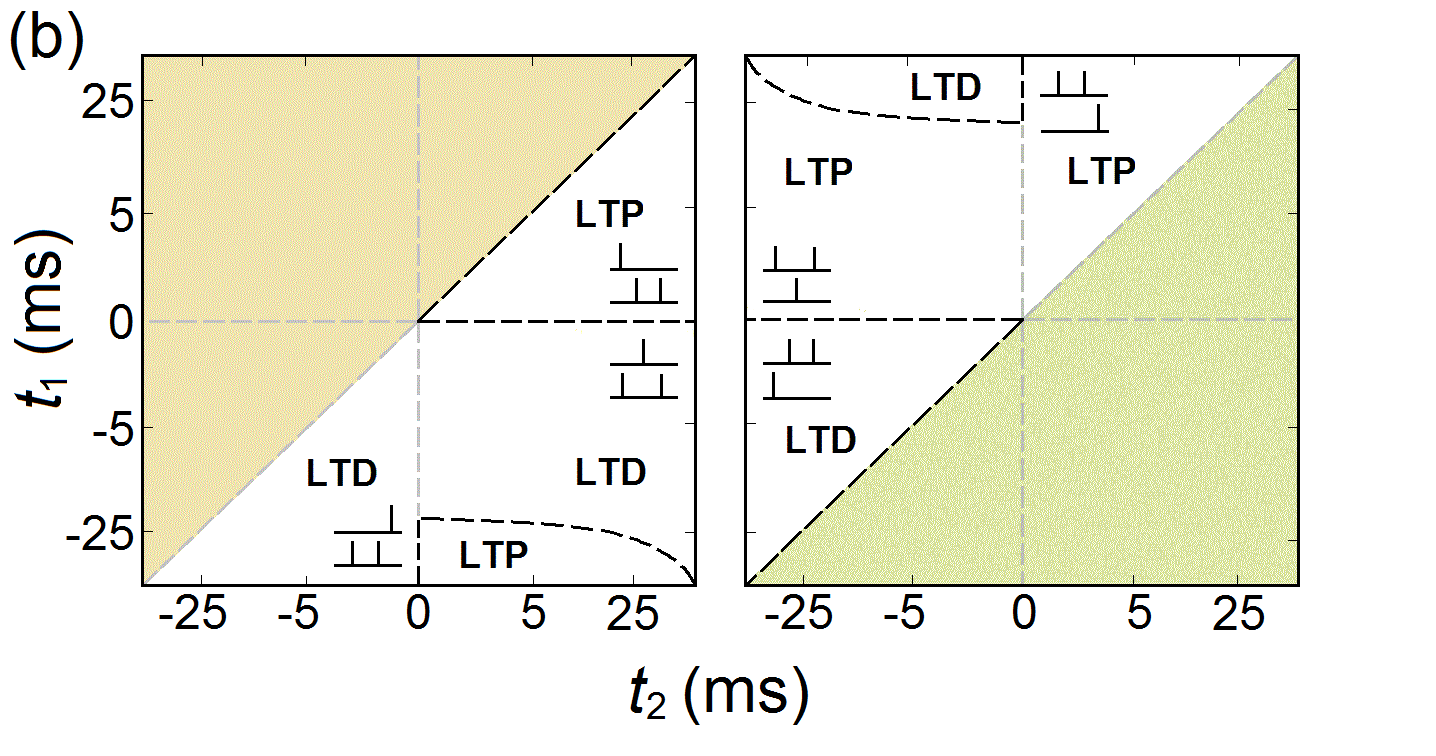}
\caption{Synaptic modifications for triplet spikes from experimental records and prediction of models. (a) Froemke's experimental data and prediction from the suppression model for '1/2' (left) and '2/1' (right) triplets. Red symbols: potentiation; blue symbols: depression. Circles: normal ACSF; triangles: high divalent ACSF with bicuculline. Dashed lines are the borders between potentiation and depression predicted by the suppression model. (b) The regions for potentiation and depression predicted by the memristive model with adaptive thresholds for '1/2' (left) and '2/1' (right) triplets. Dashed lines are the borders between LTD and LTP by the model. All the axes are plotted in logrithmic scale. The parameters are chosen as in Fig. \ref{syn}.
}\label{froe} 
\end{figure}

For a triplet of spikes, e.g., the pre-post-pre case, the first pair (pre-post) gives a positive extra voltage $\delta v^{LTP}(t)\equiv\Delta v(t)-v_{th}^{LTP}(t)$, which produces LTP. This LTP reduces the LTD threshold $v_{th}^{LTD}(t)$, so that the LTD caused by the second pair (post-pre) is suppressed by the decreased threshold. Hence, the LTP caused by the first pair is dominant in synaptic modification for triplet spikes. A numerical simulation result is plotted in Fig. \ref{syn} for typical cases (Runge-Kutta method in MAPLE v.13). An extreme case occurs when the decreased threshold $v_{th}^{LTD}(t)$ is beyond $\Delta v(t)$, which produces completely no LTD. Similar process can also be applied to the post-pre-post case, causing LTD dominantly. The net potentiation and depression regions are plotted in Fig. \ref{froe} by numerical simulations with chosen parameters, which is in accordance with Froemke's experimental records and the prediction by the suppression model. The dashed lines are boundary between LTP and LTD, corresponding to null synaptic modification or geometrically, cancelling of LTP area by LTD area. We can see that only when $t_1$ is sufficiently large, the net synaptic modification can then be predicted by the second pair, because the LTD threshold decreases to a too low level to suppress the following LTD. The boundary at large $t_1$ and $t_2$ approaches the asymptotic line with a slope of $-1$, because the LTP and LTD are independently additive, out of the time range of suppression.

Let us estimate the suppression level on the second pair quantitatively (taking the pre-post-pre case for example). As mentioned, the change of the LTD threshold caused by the LTP, denoted as $\Delta v_{th}^{LTD}$, is approximately proportional to the change of the synaptic weight $\Delta w$, which is a function of the time interval of the first pair $t_1$. If the piecewise exponential spike function $spk(t)$ is adopted \cite{Lina3}
\begin{eqnarray}\label{spk}
  spk(t) = \left\{ 
  \begin{array}{l l}
   A_{mp}^+ \frac{\textrm{e}^{t/\tau_p}-\textrm{e}^{-t_{ail}^+/\tau_p}}{1-\textrm{e}^{-t_{ail}^+/\tau_p}} &  \text{if } -t_p < t < 0\\
   -A_{mp}^- \frac{\textrm{e}^{-t/\tau_m}-\textrm{e}^{-t_{ail}^-/\tau_m}}{1-\textrm{e}^{-t_{ail}^-/\tau_m}} &  \text{if }  0 < t < t_m \\
   0 &  \text{otherwise}
  \end{array} \right.
\end{eqnarray}
 the STDP update function $\xi(t_1)=\Delta w(t_1)$ can be estimated in the piecewise exponential form
\begin{eqnarray}\label{xi}
  \xi(\Delta t) = \left\{ 
  \begin{array}{l l}
   a^+ \textrm{e}^{-\Delta t/\tau^+} &  \text{if } \Delta t>0\\
   -a^- \textrm{e}^{\Delta t/\tau^-} &  \text{if } \Delta t<0\\
  \end{array} \right.
\end{eqnarray}
Thus the total change of LTD threshold integrated during the LTP is proportional to $t_1$ exponentially
\begin{equation}\label{D_vth_D}
\Delta v_{th}^{LTD} \propto \Delta w = a^+ \textrm{e}^{-t_1/\tau^+}
\end{equation}
Thereafter, the LTD threshold increases exponentially to the resting value, described by
\begin{equation}\label{vth_D}
\begin{split}
v_{th}^{LTD}(\Delta t) -v_{th0} &=\Delta v_{th}^{LTD} \textrm{e}^{-\Delta t/\tau_{th}^{LTD}}\\
&\propto \textrm{e}^{-(t_1/\tau^+ +\Delta t/\tau_{th}^{LTD})}
\end{split}
\end{equation}
Suppose that the nonzero part of the extra voltage $\delta v^{LTD}(t)\equiv \Delta v(t)-v_{th}^{LTD}(t)$ is narrow, the amount of LTD is proportional to the extra voltage at $t_2$
\begin{equation}\label{Dw_LTD}
\Delta w^{LTD} \propto \delta v^{LTD} (t_2)= (\Delta v(t)+v_{th0})-S(t_1, t_2)
\end{equation}
with
\begin{equation}\label{S}
S(t_1, t_2)=S_p\textrm{e}^{-(t_1/\tau^+ +t_2/\tau_{th}^{LTD})}
\end{equation}
The term in the parenthesis on the rhs of Eq. (\ref{Dw_LTD}) is just the amount of LTD corresponding to the pair-based model, and the second term $S(t_1,t_2)$ gives the suppression on the LTD. We can see that the suppression amount is determined by both $t_1$ and $t_2$, which is a very wanted property. 

It is notable that this exponential dependence on the two time intervals also agrees with Froemke's suppression model. The model assumes a 'spike efficacy' of the \textit{i}th spike (supposed to be a $\delta$ function) as $\epsilon_i=1-e^{-(t_i-t_{i-1})/ \tau_s}$, which is suppressed by the preceding spike in the same neuron at time $t_{i-1}$. The synaptic modification is predicted by the pair rule multiplied by the spike efficacies $\Delta w_{ij}=\epsilon_i^{pre}\epsilon_j^{post} F(\Delta t_{ij})$, where $\Delta t_{ij}=t_j^{post}-t_i^{pre}$ is the time interval betwen the post- and pre-synaptic spikes and function $F$ is defined as
\begin{eqnarray}\label{F}
  F(\Delta t) = \left\{ 
  \begin{array}{l l}
   A^+ \textrm{e}^{-\Delta t/\tau_+} &  \text{if } \Delta t>0\\
   -A^- \textrm{e}^{\Delta t/\tau_-} &  \text{if } \Delta t<0\\
  \end{array} \right.
\end{eqnarray}
By these definitions, we can derive the suppression on the LTD by the preceding LTP to be $-A^- e^{-(t_1/\tau_s + t_2/\tau_s')}$, with $\tau_s'\equiv\tau_s\tau_-/(\tau_- -\tau_s)$. This exponential dependence on $t_1$ and $t_2$ is accordant with the previously derived suppression function $S(t_1,t_2)$ in the memristive model. Hence, the memristive model with variable thresholds can be regarded as another expression of the suppression principle for the triplet STDP rule. However, the notable difference from the suppression model is that the memristive model imposes suppression on the following LTD or LTP, instead of directly on the spikes.

\section{Generalized Form and Biological Relations}
The memristive model in Eq. (\ref{dvth_sim1}) and (\ref{dvth_sim2}) can be regarded as a special form of the general model in the following form, which assumes to adapt both two thresholds to the presence of LTP or LTD
\small\begin{eqnarray}\label{dvth_ful}
 \frac{dv_{th}^{LTP}}{d t}=\frac 1 {\tau_{th}^{LTP}} \left(-v_{th}^{LTP}+v_{th0}\right)+k_{PP}\left[\frac{dw}{dt}\right]_- + k_{PD}\left[-\frac{dw}{dt}\right]_+\nonumber\\
\\
 \frac{dv_{th}^{LTD}}{d t}=\frac 1 {\tau_{th}^{LTD}} \left(-v_{th}^{LTD}-v_{th0}\right)+k_{DP}\left[\frac{dw}{dt}\right]_- + k_{DD}\left[-\frac{dw}{dt}\right]_+\nonumber\\
\end{eqnarray} \normalsize
where the term with $k_{PP}$ (positive) indicates that the presence of LTP causes decrease in the LTP threshold, while the term with $k_{DD}$ (positive) indicates that the presence of LTD causes increase in the LTD threshold. Hence, LTP will enhance the following LTP (also the current LTP itself) and suppress the following LTD; and vice versa. The terms with factors $k_{PD}$ and $k_{DP}$ impose suppression as previous. We are not intended to identify the specific biological quantities for the variable thresholds in this mechanistic model, but certain relations may be addressed. The variable thresholds can be viewed as an expression of short term plasticity. Since STDP depends on NMDA receptor activation or glutamate bound for presynaptic events, and on the rise in the voltage-dependent influx Ca$^{2+}$ concentration level and NMDA channels for postsynaptic events, the variable thresholds are correlated with the probability of release of neurotransimitters $P_{rel}$ and the probability of the postsynaptic channel openning $P_{s}$, both of which reveal short term plasticity by exhibiting a rapid rise to a high level and returning back exponentially to the resting value \cite{Abbo}. In addition to chemical nature, synapses with gap junctions may be regarded as a more direct correspondance to the memristive model \cite{Abbo}\cite{Kand}, which produce a synaptic current proportional to the difference between the pre- and post-synaptic membrane potentials. In this case, the variable thresholds can be determined directly by inner electric properties of the synapse.    

In addition, we want to point out that this memristive STDP model can also be viewed as a rate-based model, regarding $v_{pre}(t)$ and $v_{post}(t)$ as the rate function of pre- and post-synaptic events, instead of the membrane potentials. Therefore, in general, the equation of synaptic modification rate can be written as a general function on the excessive rate difference over the thresholds
\begin{equation}\label{rate_1}
\frac{d\textbf{w}}{d t}=\textrm{G}\left( [\textbf{F}(t)-\textbf{F}_{th}]_+ , [\textbf{F}(t)+\textbf{F}_{th}]_-\right)
\end{equation} 
with
\begin{equation}\label{rate_2}
\textbf{F}(t)=(v_{post}(t)-v_{post0})\textbf{n}-(\textbf{v}_{pre}(t)-\textbf{v}_{pre0})
\end{equation}
where the vector defines the rate functions, referred to the resting rates $v_{post0}$ ($\textbf{n}$ is the unit vector) and $\textbf{v}_{pre0}$, corresponding to multi-presynaptic neurons (connected to one postsynaptic neuron), among which competition can be introduced.

\section{Conclusions}
As we have demonstrated, the new mechanism of adaptive thresholds is an expression of the principle of suppression, which can make the memristive STDP model compatible with the both pair and triplet rules. The exponential dependence on the time intervals is comparable with that predicted in the suppression model. The net synaptic potentiation and depression regions are also accordant with the experimental results and the prediction by the suppression model. Possible biological quantities related to the adaptive thresholds are addressed, which suggests an expression of short term plasticity. However, the most important is that we can foresee that this model may be used by neuromorphic engineers in the realization of emulators of biological cortical networks, for which implementations of suitable memristor devices or equivalent circuit designs are demanded.

\ifCLASSOPTIONcaptionsoff
  \newpage
\fi


%\vfill


\end{document}